\documentclass[conference]{IEEEtran}
\IEEEoverridecommandlockouts

\usepackage{cite}
\usepackage{amsmath,amssymb,amsfonts}
\usepackage{graphicx}
\usepackage{textcomp}
\usepackage{xcolor}
\usepackage{array}
\usepackage{booktabs}
\usepackage{tikz}
\usepackage{pgfplots}
\pgfplotsset{compat=1.18}
\usetikzlibrary{arrows.meta, shapes.geometric, positioning, fit, backgrounds, decorations.pathreplacing}

\def\BibTeX{{\rm B\kern-.05em{\sc i\kern-.025em b}\kern-.08em
    T\kern-.1667em\lower.7ex\hbox{E}\kern-.125emX}}

\definecolor{teal}     {RGB}{10,  126, 164}
\definecolor{navy}     {RGB}{13,  27,  42}
\definecolor{amber}    {RGB}{212, 133,  10}
\definecolor{green}    {RGB}{26,  113,  90}
\definecolor{purple}   {RGB}{91,   63, 160}
\definecolor{graylt}   {RGB}{230, 235, 240}
\definecolor{corallt} {RGB}{253, 234, 234}

\usepackage{soul}

\begin{document}

\title{Human-Centricity in Industry 5.0: \\ A Survey of Worker Sensing, Adaptive Operations, and Human-in-the-Loop Systems
%
}


\author{\IEEEauthorblockN{
Lara Pereira\IEEEauthorrefmark{1},
Luís Miguel D. F. Ferreira\IEEEauthorrefmark{2} and
João Ruivo Paulo\IEEEauthorrefmark{1}}

\IEEEauthorblockA{
\IEEEauthorrefmark{1}Institute of Systems and Robotics, University of Coimbra, Portugal\\
Emails: \{lara.pereira, jpaulo\}@isr.uc.pt}

\IEEEauthorblockA{
\IEEEauthorrefmark{2}Centre for Mechanical Engineering, Material and Processes, University of Coimbra, Portugal\\
Email: luis.ferreira@dem.uc.pt}
}

\maketitle

\begin{abstract}
Industry~5.0 (I5.0) repositions manufacturing around a human-centric vision in which cyber-physical systems must adapt to the worker rather than the other way around.
Despite significant advances in worker state monitoring
technologies, including markerless computer vision, wearable
physiological sensors, and machine-learning-based fatigue and cognitive load estimation, a critical integration gap persists: these monitoring capabilities rarely inform real-time operational decisions, and workers almost never participate in the adaptive processes that affect their
tasks and workloads.
This paper presents a structured survey of the state-of-the-art across three interconnected dimensions: (1)~worker state monitoring in industrial settings, encompassing physical fatigue, cognitive load, and ergonomic risk assessment; (2)~human factors integration in Operations
Management (OM) models, examining how scheduling, task allocation, and production planning incorporate worker-related objectives; and (3)~Human-in-the-Loop (HITL) architectures, analyzing the extent to which worker feedback closes the adaptive decision loop.
The survey reveals that while each dimension has matured independently, the integration across them remains largely absent.
Based on the reviewed literature, no publicly documented and validated system connects real-time sensor-driven worker state estimation with adaptive OM decisions and worker feedback in a unified closed-loop architecture.
This gap is identified as the central challenge for realizing the I5.0 vision, and a five-direction research agenda is proposed.
\end{abstract}


\section{Introduction}

Occupational fatigue and musculoskeletal disorders represent a critical vulnerability in modern manufacturing, driving severe human and economic costs through lost productivity and workplace injury~\cite{Choobineh2026}. Despite this, Operations Management (OM) frameworks governing daily production, such as scheduling, task allocation, and workflow control, operate as if worker physiology were constant. Historically engineered for throughput maximization and cycle-time reduction, traditional OM models reduce human operators to standardized resources with fixed service rates~\cite{Prunet2024}, a perspective we term the worker-as-resource assumption, in which physiological and cognitive variability is treated as noise rather than a state variable to be managed. By ignoring the dynamic fluctuations in cumulative fatigue, stress, and cognitive overload, these models assign identical workloads regardless of individual capacity. Overcoming this worker-as-resource assumption requires fully realizing the vision of Industry 5.0 (I5.0): deploying cyber-physical systems that actively sense, model, and dynamically adapt to the human worker in real time.

Currently, real-time monitoring of worker state has achieved significant technical maturity. Wearable sensor networks can continuously capture physiological telemetry, including heart rate variability (HRV), electrodermal activity (EDA), and electromyography (EMG)~\cite{Mohapatra2024}. Moreover, markerless computer vision (CV) frameworks extract 3D kinematics and compute ergonomic risk scores without requiring intrusive instrumentation~\cite{Agostinelli2024}. Leveraging these signals, modern machine learning (ML) architectures can classify and predict muscular fatigue, cognitive load, and ergonomic risk with clinically relevant accuracy.

Despite this technological readiness, a severe integration gap persists. While the algorithms required to adapt OM decisions are computationally feasible, they remain largely disconnected from this live physiological telemetry. Partial integration attempts exist, yet they are highly fragmented. For example, platforms like RT-PROFASY~\cite{Donati2022} aggregate physiological and production data for retrospective managerial analysis but fail to trigger automated operational adaptations. Conversely, adaptive human-robot interaction (HRI) systems successfully respond to cognitive load in real time~\cite{Hostettler2024}, but operate strictly at the level of localized robotic behavior rather than governing shop-floor scheduling or task allocation.

Furthermore, while conceptual I5.0 frameworks advocate for this closed-loop integration, they remain largely theoretical, with literature explicitly acknowledging a persistent lack of empirical validation~\cite{Vido2025}. Even sophisticated adaptive controllers, such as the CEPC~\cite{Wong2022}, which enable real-time workflow adjustments and human override, rely exclusively on machine and workpiece states, meaning the system does not incorporate human physiological or cognitive telemetry. Ultimately, no validated closed-loop architecture has been identified that connects live, sensor-derived worker state estimates directly to OM-level scheduling decisions, nor does any system incorporate active worker feedback to modulate these automated adaptations in an industrial setting~\cite{Prunet2024}.

To explore this disconnect and provide a baseline for cross-disciplinary discussion, this paper reviews the technical and systemic boundaries separating these research fields. Within this context, the contributions of this survey are threefold:
(1)~a structured synthesis of the state-of-the-art across three traditionally disparate research domains; 
(2)~an empirical mapping of the integration gap separating biometric monitoring, OM adaptation, and active worker feedback; and 
(3)~a strategic research agenda detailing the specific technical and methodological challenges that must be resolved to operationalize the human-centric mandate of I5.0.

\section{Methodology and Scope}
This survey adopts an umbrella-review approach (a synthesis of existing systematic and scoping reviews), supplemented by primary experimental literature and open-source datasets where review-level evidence is still emerging. Literature was identified across five databases - IEEE Xplore, Scopus, Web of Science, ScienceDirect, and the ACM Digital Library - covering peer-reviewed publications from 2018 to 2026, using the boolean string:

\begin{quote}
\textit{("Worker State" OR "Ergonomic Risk" OR "Cognitive Load" OR "Fatigue") AND ("Operations Management" OR "Scheduling" OR "Task Allocation") AND ("Human-in-the-Loop" OR "Adaptive" OR "Feedback") AND ("Manufacturing" OR "Industry 5.0" OR "Shop Floor" OR "Cyber-Physical System")}
\end{quote}

Retrieved literature was categorized into four thematic dimensions - {Perception and Sensing (fatigue, ergonomic, and cognitive-load estimation methods), Operational Optimization (scheduling and task-allocation models incorporating human factors), Explainability (translating optimization outputs into operator-facing instructions), and Interaction and Agency (architectures for worker override, negotiation, and trust calibration) - which map directly onto the four architectural layers proposed in Section VII (Sense, Decide, Explain, Feedback).

Inclusion required peer-reviewed, English-language studies explicitly modeling the interface between operator states and industrial or logistics workflows; studies confined to sedentary office settings, standalone sensor/algorithm validation without systems-level implications, or unvalidated out-of-domain laboratory work were excluded. Foundational datasets and technical benchmarks cited within included studies (e.g., pose-estimation or physiological-signal datasets) were retained as supporting references despite not themselves modeling the operator-OM interface. This process yielded a core pool of 39 papers. Our analysis of the integration gap is limited to this peer-reviewed, open literature; proprietary industrial solutions and grey literature fall outside its scope.

\section{Industry~5.0 Paradigm}

While Industry~4.0 (I4.0) established a robust technological foundation centered on automation, Internet of Things (IoT), and cyber-physical systems (CPS) to optimize operational efficiency, I5.0 reorients its purpose around three pillars: sustainability, resilience, and human-centricity~\cite{EC_ERA_Roadmap_2024, Vido2025}. 
Specifically, within this human-centric pillar, on both technological and organizational levels, the focus is on placing human needs, characteristics, motivation, and experiences at the center of design, within a transition that emphasizes enhanced human-machine collaboration, aiming for human physical and cognitive well-being, dynamic adaptation to individual capacities, and respect for worker agency~\cite{EC_ERA_Roadmap_2024}.

Building upon the technological foundations of I4.0, advanced CPS, including robotics and artificial intelligence (AI), are repurposed to cultivate a flexible and resilient synergy between human operators and modern manufacturing systems~\cite{Vido2025, Clemmensen2025}.
To achieve this, these platforms must transcend traditional machine-to-machine coordination, evolving into HITL architectures that should continuously sense, model, and dynamically respond to human cognitive and physical states~\cite{Clemmensen2025}. 
For this, at the perception layer, sensors must capture physiological and kinematic state; at the intelligence layer, models should interpret that state in real time; at the actuation layer, decisions adapt workflows and task assignments; and at the interaction layer, workers must not only understand and influence system adaptations but must be able to calibrate their trust in those adaptations over time.

The European Commission's ERA technology roadmap on human-centric approaches~\cite{EC_ERA_Roadmap_2024} identifies six categories of enabling technologies for I5.0, covering tracking technologies for mental and physical strain, improving cognitive human capabilities through AI and decision support systems, and adaptive automation with trust calibration~\cite{EC_ERA_Roadmap_2024}. 
Consequently, these categories directly align with the four-layer sense–decide–explain–feedback architecture proposed later in this survey (Section VII), suggesting that a sensing-to-feedback stack reflects an institutional policy requirement rather than a purely theoretical construct.
Existing reviews of I5.0 in manufacturing find that psychosocial dimensions, including worker autonomy, trust, motivation, and stress are significantly underrepresented, and a disconnect persists between the theoretical frameworks proposed and their empirical validation~\cite{Passalacqua2025,Vido2025}.

The present survey addresses this gap by specifically examining the integration of worker state monitoring and operational decision-making as the concrete mechanism through which I5.0's human-centric vision must be realized. Taken together, the ERA roadmap's enabling-technology categories and the four-layer sense–decide–explain–feedback architecture proposed to operationalize them (Section VII) define the scope of this survey, where Section IV examines the sensing layer, Section V the decision layer, and Section VI the explain and feedback layers.

\section{Worker State Monitoring in Industrial Settings}
\label{sec:sensing}
\subsection{Physical Fatigue and Ergonomic Risk}

The European Commission's ERA technology roadmap for I5.0~\cite{EC_ERA_Roadmap_2024} establishes worker state monitoring not merely as an academic pursuit, but as a mandatory operational capability. Specifically, the roadmap identifies the continuous monitoring of physical and cognitive strain as a key enabling technology for human-centric manufacturing~\cite{EC_ERA_Roadmap_2024}.
Assessing the risks of work-related musculoskeletal disorders (WMSDs) and ensuring the well-being, productivity, and safety of the operator should be prioritized, with physical fatigue and musculoskeletal ergonomic assessments as essential dimensions of monitoring the worker's condition in industrial settings~\cite{Scataglini2025, Mohapatra2024}. 

Traditional methods rely on observational techniques administered by specialists, such as Rapid Whole Body Assessment (REBA) and Rapid Upper Limb Assessment (RULA), which provide validated risk scores but require trained assessors, are susceptible to subjectivity, and cannot be applied continuously~\cite{Daz2026, Garca-Luna2024}. Driven by advancements in wearable kinematic sensors and CV, ergonomic assessment has experienced a decisive transition from these methods to automated, continuous monitoring. A recent scoping review of 84 studies confirms this shift, yet cautions that a persistent lack of standardization in validation protocols continues to restrict the generalizability of these sensor-based approaches across industrial settings~\cite{Iyer2025}.

Exemplifying the deployment of wearable systems, Mohapatra et al.~\cite{Mohapatra2024} demonstrated a multimodal sensor network, combining ECG, HRV, skin temperature, and IMUs that achieved continuous, multilevel physical fatigue prediction across 43 workers in a real manufacturing environment. However, their findings exposed an important vulnerability: subject-dependency. Models trained on one demographic group suffered significant performance degradation when generalized, establishing demographic-aware AI as an unresolved prerequisite for scalable deployment~\cite{Mohapatra2024}.
Moreover, sensor- and marker-based motion capture systems are inherently invasive and require precise calibration by trained personnel. These setups can physically hinder natural movement and inadvertently alter the operator’s normal behavior ~\cite{Baklouti2024, Agostinelli2024}. Consequently, AI-based markerless CV has emerged as a compelling, non-intrusive alternative. 
By leveraging deep learning models, such as OpenPose, MediaPipe, and Convolutional Pose Machines (CPM) like Smart Ergonomic Explorer (SEE), these systems can extract 2D skeleton keypoints using just one or two standard RGB cameras or smartphones, and triangulate  them into three-dimensional joint angles using open-source multi-view pipelines such as Pose2Sim~\cite{Pagnon2022}. This eliminates sensor overhead, offering high portability and flexibility without the need for constant recalibration~\cite{Agostinelli2024}.

Currently, researchers are bypassing explicit skeleton fitting by leveraging large-scale human motion datasets. Using benchmarks like Human3.6M~\cite{Ionescu2014}, MPI-INF-3DHP~\cite{Mehta2016}, or custom industrial RGB-D datasets~\cite{Sapoutzoglou2026}, modern deep learning frameworks directly map raw video feeds to ergonomic metrics. This facilitates automated, frame-level RULA scoring, eliminating the computational overhead of manual joint-angle calculation and the need for invasive worker instrumentation.

However, a key validation gap remains: the leap from laboratory to factory floor. Despite algorithmic advances, validation occurs predominantly under controlled conditions. The literature lacks systematic assessments in real-world manufacturing environments marked by severe occlusions, dynamic lighting, and unpredictable operator mobility. Furthermore, existing validation is often misdirected, focusing on joint-angle precision rather than the reliability of the resulting ergonomic risk scores.
Even the most advanced real-world deployments are functionally static. Menanno et al.~\cite{Menanno2024}, for instance, used pre-recorded RGB video to identify high-risk tasks and reassign them to cobots, reducing ergonomic stress by 13\%. However, this system tracks a single offline operator to produce fixed workstation redesigns. It cannot dynamically adapt task allocation to a worker's real-time fatigue state. This limitation underscores the central challenge identified by this survey: bridging the gap between static ergonomic alerting and live, adaptive OM integration.

\subsection{Cognitive Load and Mental Fatigue}

Cognitive load estimation in industrial settings is substantially less developed than physical fatigue monitoring.
A 2025 systematic literature review of cognitive workload (CWL) assessment methods in the I5.0 context confirms that there is still no reliable methodology to assess CWL in different work environments, and that practical methodologies for real-time monitoring and predictive modeling are critically lacking~\cite{Vitti2025}.
Standard approaches to assessing CWL typically rely on basic performance outputs (e.g., completion times, error rates) or subjective self-reporting (e.g., NASA-TLX, SMEQ)~\cite{Mndez2025,Lucchese2025}. However, both methods are poorly suited for real-time systems. Performance metrics fail to isolate cognitive strain from other operational variables, while subjective surveys necessitate task interruption and suffer from severe retrospective bias and memory distortion. Consequently, these ex-post evaluations are operationally subjective and incapable of driving the live, adaptive OM decisions required by Layer 2 of the proposed framework (Section \ref{sec:gap}).

To achieve continuous, objective measurements, research has prioritized direct neural correlates via electroencephalography (EEG), eye-tracking~\cite{Vitti2025, Lucchese2025}, and functional near-infrared spectroscopy (fNIRS)~\cite{Han2023}. While these modalities offer high temporal resolution and localized cortical mapping, their industrial viability is compromised by the need for specialized hardware, sensor intrusiveness, and acute susceptibility to motion artifacts~\cite{Mndez2025}. 
To balance precision with operator mobility, the field has increasingly pivoted to minimally invasive peripheral signals, utilizing HRV and EDA as reliable CWL proxies~\cite{Rezaei2025}. 
However, deploying these metrics into actionable systems is significantly limited by a scarcity of operationally valid data. Existing open-source datasets broadly fail to represent industrial realities. Representative examples are either restricted to minor laboratory studies (e.g., WESAD~\cite{Schmidt2018}) or target stationary, out-of-domain tasks (e.g., ADABase~\cite{Oppelt2022}). Furthermore, the widespread reliance on ex-post questionnaires across these types of datasets introduces severe retrospective bias. Even as emerging multimodal datasets (MultiPhysio-HRC~\cite{Bussolan2025}) attempt to address industrial settings, they remain constrained by simulated environments and coarse labels.

Crucially, the literature lacks a unified dataset capable of simultaneously tracking physical fatigue, cognitive load, and ergonomic risk in uncontrolled manufacturing environments.
To bypass both the physical intrusiveness of contact sensors and the severe limitations of existing datasets, the field is shifting toward observable kinematics. Specifically, an emerging research direction leverages markerless CV as a non-intrusive kinematic proxy for cognitive load. By dispensing with wearable instrumentation entirely, this approach ensures scalable deployment across heterogeneous worker populations. Structurally, frameworks like $\mathrm{T}^2\mathrm{W-CogLoadNet}$~\cite{Zhao2026} have demonstrated that subtle degradations in movement quality, reach velocity, and postural stability can map directly to mental workload under controlled conditions. While the ergonomics community has independently prioritized the integration of physical and cognitive factors through combined motion tracking \cite{Iyer2025}, this vision-based kinematic-cognitive proxy has yet to be validated in real manufacturing environments. 
Closing this validation gap is one of the central objectives of Direction 1 of the research agenda proposed in Section \ref{sec:gap}.

\section{Human Factors in Operations Management}
\label{sec:om}

\subsection{The Worker-as-Resource Assumption}

Conventional OM paradigms prioritize system throughput and profit, reducing human workers to homogeneous standardized resources~\cite{Prunet2024}. Traditional optimization frameworks (e.g., Mixed-Integer Linear Programming (MILP) and metaheuristics) restrictively assume time-invariant human productivity, systematically flattening human variability into static efficiency coefficients rather than treating it as a dynamic physiological state~\cite{Gao2025}. Despite the real-time adaptive capabilities of I4.0 technologies like IoT and digital twins, current scheduling algorithms remain paralyzed by these rigid models~\cite{Prunet2024}.
This architectural oversight yields significant operational inefficiencies. Thürer et al.~\cite{Threr2020} demonstrated that ignoring fatigue dynamics, specifically the inevitable performance crash that follows high-load acceleration, directly deteriorates overall shop-floor efficiency. Furthermore, Grosse et al.~\cite{Grosse2023} warn that the financial gains projected by traditional efficiency-focused models are largely "phantom profits" quickly erased by the costly errors, dysfunction, and absenteeism caused by unmanaged human strain. 
Consequently, relying on static-resource assumptions  limits the operational resilience of modern manufacturing, rendering theoretical schedules obsolete the moment operator fatigue deviates from the mathematical mean.

\subsection{Human-Aware Scheduling Models}

While OM research increasingly acknowledges human factors, its current integration remains strictly analytical and fundamentally disconnected from the worker's live physical state. Prunet et al.~\cite{Prunet2024} extensively catalog human-aware scheduling models, revealing a systemic limitation: variables such as fatigue, learning curves, and ergonomic exposure are overwhelmingly reduced to static thresholds or aggregated population averages. They are never captured as live, sensor-derived estimates that evolve dynamically during a shift~\cite{Prunet2024}. Similarly, while Grosse et al.~\cite{Grosse2023} formalize theoretical OM constructs for physiological state indicators, these remain abstract mathematical assumptions isolated from real-time biometric tracking.

Even explicit I5.0 frameworks fail to bridge this integration gap. The Human-Centric Shop-Floor Scheduling (HCSFS) paradigm advocates for an equal balance between production efficiency and worker well-being~\cite{Gao2025}. Yet, literature remains stubbornly "human-related" rather than truly "human-centric", treating workers merely as static constraints rather than dynamic, active decision-makers. In Human-Robot Collaboration (HRC), systems by Baratta et al.~\cite{Baratta2024} dynamically allocate tasks but still rely on a pre-defined, exponential fatigue curve rather than actual physiological feedback. Furthermore, sophisticated adaptive controllers such as the CEPC~\cite{Wong2022} dynamically adjust workflows based on machine states but remain completely blind to the human operator~\cite{Wong2022}. Exacerbating this issue, algorithmic task allocation demands worker trust calibration, an essential I5.0 design requirement that is ignored by purely analytical models~\cite{EC_ERA_Roadmap_2024}. 
While these mathematical formulations successfully prove that operator fatigue can be optimized, their reliance on generalized, offline decay equations restricts their practical execution in highly dynamic production environments.

\subsection{Ergonomic Risk in Scheduling}

This reliance on static proxies is particularly evident in ergonomic optimization. While integrating metrics like RULA and REBA into scheduling is an I5.0 cornerstone~\cite{Iyer2025}, current methodologies misattribute this risk to the workstation rather than the worker. Modern MILP and HRC formulations successfully elevate well-being to a primary objective, co-optimizing makespan alongside ergonomic exposure and cognitive load. However, because these scores are permanently tied to pre-classified operations during the design phase, they inherently ignore individual biomechanics and accumulating physical exposure. Consequently, even state-of-the-art industrial vision systems function purely as diagnostic tools, generating indices solely for retrospective task redesign rather than driving live operational adaptations~\cite{Menanno2024,Iyer2025}. 
As the ERA technology roadmap explicitly prioritizes adaptive automation~\cite{EC_ERA_Roadmap_2024}, the operational focus must now pivot from passively quantifying ergonomic constraints to utilizing these metrics as live, dynamic triggers for workflow adaptation.

\section{Human-in-the-Loop Architectures in Industrial Systems}
\label{sec:hitl}

\subsection{HITL Taxonomy and Scope}

A critical sociotechnical baseline for this paradigm is established by Clemmensen et al.~\cite{Clemmensen2025}. Through a systematic review of 60 papers paired with expert interviews among industry practitioners, the authors identify four recurring knowledge domains essential to cyber-physical integration: interface design, adaptive system architecture, organizational context, and design methodology. Crucially, their findings expose systematic discrepancies between how academic literature conceptualizes trust, oversight, and control, and how industrial experts actually experience them~\cite{Clemmensen2025}. Arrived at independently of mainstream manufacturing frameworks, this insight directly corroborates that theoretical HITL modeling has significantly outpaced empirical, field-validated integration.

In modern industrial CPS, the HITL paradigm extends beyond passive system oversight. It elevates workers to active agents, embedding human cognition in technical systems to prevent cognitive overload rather than automate human judgment away~\cite{Emmanouilidis2019,Turner2021}. Emmanouilidis et al.~\cite{Emmanouilidis2019} define this as a supervisory loop for situational awareness: the technical system contextualizes complex data and recommends actions, but the human operator retains the ultimate authority to evaluate and execute. This synergy is bidirectional. Empowered by natural language interfaces and visual analytics, human operators make faster, more informed decisions, while their iterative feedback simultaneously trains and refines ML diagnostic models~\cite{Emmanouilidis2019}. While the theoretical scope of this integration spans the entire manufacturing stack, from shop-floor operations to enterprise ERP and SCADA systems~\cite{Emmanouilidis2019}, practical implementation is still open. Precise alignment of these bidirectional loops with human cognitive limits to drive adaptive OM decisions in real-time remains an open research gap~\cite{Emmanouilidis2019} that this survey addresses.

\subsection{Trust Calibration and Worker Agency}

The success of industrial HITL systems hinges on effective trust calibration: preventing workers from blindly accepting (overtrust) or systematically rejecting (undertrust) AI decisions. The CalTruIAS model~\cite{Lucas2024} maps this dynamic across three evolving dimensions of trust: benevolence, competence, and integrity. Because these dimensions fluctuate constantly during human-machine interaction, static interfaces are poorly suited to manage them. Instead, recent cognitive modeling demonstrates that active system feedback is paramount; systems must continuously broadcast their confidence levels and operational limitations in real time. However, merely communicating uncertainty is useless if the operator lacks the authority to act upon it. Consequently, worker agency, the ability to override or negotiate decisions, is a core I5.0 mandate, not an optional feature~\cite{Passalacqua2025,EC_ERA_Roadmap_2024}.
If physiological monitoring drives task adaptation without transparency, it can trigger significant institutional distrust regarding job security and privacy, primary drivers of undertrust in workplace AI~\cite{Lucas2024}. Layer 4’s architectural imperative is clear: design lightweight, highly transparent feedback mechanisms that sustain calibrated trust and genuine autonomy across an entire shift.

This challenge escalates dramatically when the adaptive agent is a physically co-present humanoid robot rather than a disembodied algorithm. Interestingly, humanoid appearance fosters more stable trust dynamics; workers apply the forgiving expectations reserved for human colleagues rather than the rigid perfection expected of machines~\cite{Lucas2024}. However, humanoids introduce complex dimensions absent from standard models: anthropomorphism, perceived agency, physical proximity, and non-verbal communication~\cite{Lucas2024,Bajestani2025}. Worker agency here transcends overriding a workflow; it requires negotiating shared physical space and delegating tasks to a human-like entity. As humanoid platforms near deployment readiness in manufacturing~\cite{Bajestani2025,EC_ERA_Roadmap_2024}, navigating these embodied socio-technical dynamics defines the frontier for HITL research, a challenge explicitly addressed in Direction 5.

\subsection{Explainability as a HITL Prerequisite}

Meaningful worker participation in adaptive OM decisions requires that those decisions be explainable in operational rather than technical terms. Standard XAI methods, such as SHAP, LIME, and attention mechanisms, produce feature importance attributions that are interpretable to data scientists but rarely actionable for factory floor operators~\cite{Tzionis2026}.
The distinction is illustrated concretely by a state-of-the-art worker-state monitoring system recently validated in industrial settings: Mohapatra et al.~\cite{Mohapatra2024} apply SHAP to identify chest IMU angular velocity and maximum heart rate as dominant fatigue predictors, and present these through a near-real-time dashboard showing physiological signs and a continuous fatigue score. The authors identify adaptive work scheduling as the intended downstream application of these estimates~\cite{Mohapatra2024}, but the dashboard does not provide a mechanism to translate sensor output into workstation-level instructions that an operator or scheduler can act on immediately.
This gap between technically correct and operationally actionable explanations is not a limitation specific to 
Mohapatra et al., it reflects the absence of a formalized framework to translate model outputs into role-specific, context-aware instructions for factory-floor operators~\cite{EC_ERA_Roadmap_2024}.
The ERA technology roadmap explicitly identifies the need for \textit{``XAI for interactive support''} and \textit{``situation-aware assistance systems''} as open technology categories~\cite{EC_ERA_Roadmap_2024}, confirming that operational explainability remains an unresolved challenge at both the policy and technical levels.
This gap constitutes the main motivation for Direction~2 of the research agenda proposed in Section~\ref{sec:gap}.

\section{Proposed Framework and Integration Gap}
\label{sec:gap}

Based on the survey findings, Fig.~\ref{fig:framework} proposes a four-layer human-centric framework that closes the integration gap identified across Sections~\ref{sec:sensing}--\ref{sec:hitl}.
Each layer corresponds to a capability that the literature has developed
independently; the framework connects them into a continuous adaptive
loop.

\begin{figure}[t!]
  \centering
  \includegraphics[width=3.5in]{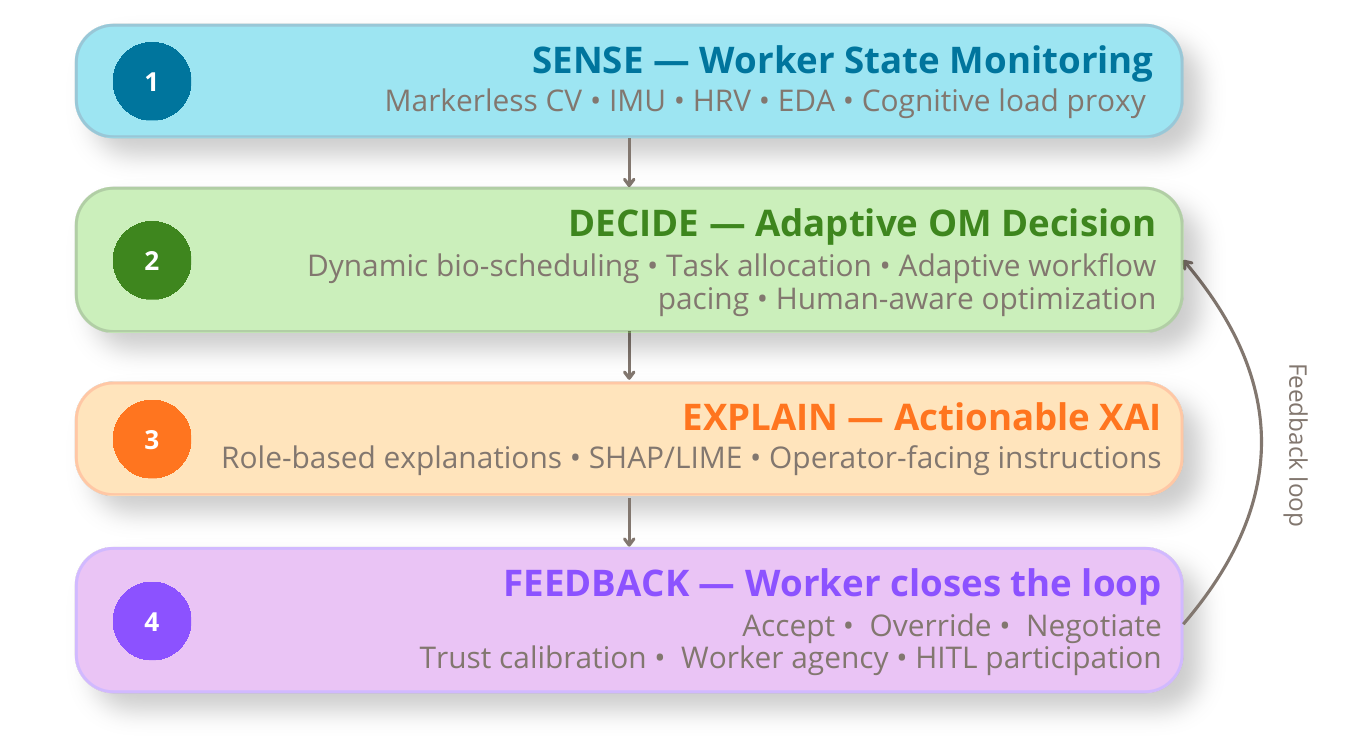}
  \caption{Proposed four-layer human-centric framework.}
  \label{fig:framework}
\end{figure}

\begin{table}[t]
\caption{Integration Status Across Surveyed Dimensions}
\label{tab:gap}
\begin{center}
\renewcommand{\arraystretch}{1.25}
\begin{tabular}{>{\raggedright\arraybackslash}p{1.2cm} 
                >{\raggedright\arraybackslash}p{1.95cm} 
                >{\raggedright\arraybackslash}p{2.3cm} 
                >{\raggedright\arraybackslash}p{1.4cm}}
\toprule
\textbf{Dimension} & \textbf{Current State} & \textbf{Key Gap} & \textbf{Layer} \\
\midrule
Worker state monitoring & Physical fatigue: mature. Cognitive load: limited industrial translation. Demographics: absent. & No system fuses physical and cognitive modalities in real factory settings. & Layer~1 (Sense)\\
OM with human factors & Static ergonomic constraints; analytical fatigue models in scheduling. & Worker state is mathematically assumed rather than sensor-driven. & Layer~2 (Decide)\\
HITL architecture & Trust calibration theory; physical safety protocols. & Active worker feedback on OM decisions is absent. & Layers 3--4  (Explain/ Feedback)\\
\midrule
\textbf{Full integration} & 
\multicolumn{2}{>{\raggedright\arraybackslash}p{4.9cm}}{\textbf{No validated system has been identified that connects live sensing $\rightarrow$ OM adaptation $\rightarrow$ worker feedback in an industrial setting.}} & 
\textbf{NOT ACHIEVED} \\
\bottomrule
\end{tabular}
\end{center}
\end{table}

The four layers were derived inductively from the gap analysis in Table I rather than imposed a priori. Layer 1 (Sense) corresponds directly to the worker-state-monitoring literature reviewed in Section IV. Layer 2 (Decide) corresponds to the OM-with-human-factors literature reviewed in Section V. The HITL literature reviewed in Section VI, by contrast, does not map onto a single layer, it repeatedly conflates two functionally distinct capabilities, making an automated decision interpretable to a worker (Section VI-C) and giving the worker the standing to accept, override, or negotiate that decision (Section VI-B).

Table~\ref{tab:gap} summarizes the integration status in all dimensions surveyed, mapping each to the corresponding framework layer.
The gap has three structural causes.
First, disciplinary siloing: worker state monitoring research is primarily conducted in biomedical engineering, human factors, and CV communities, while OM research resides in industrial engineering and management science, communities with limited cross-disciplinary dialogue. This fragmentation is quantitatively visible in the production engineering literature, where a recent systematic review found papers on human-centric I5.0 distributed across 19 different journals with no single outlet publishing more than three contributions~\cite{Vido2025}.
Second, methodological mismatch: sensor-derived worker state estimates are probabilistic, noisy, and person-specific, while OM optimizers require deterministic, structured inputs; bridging this interface requires both technical innovation and formal modeling.
Third, validation barriers: validating a closed-loop HITL-OM
system requires access to real industrial facilities,
multi-disciplinary instrumentation, and participant recruitment at scale.

\subsection{Research Agenda}
Five research directions are proposed to close the integration gap.

\textbf{Direction~1 - Multimodal Worker State Estimation with Demographic Adaptation.}
Current sensing systems capture either physical fatigue or cognitive load, rarely both simultaneously, and few account for demographic variability.
A validated pipeline combining multi-view markerless CV
with wearable physiological sensors is needed, targeting simultaneous estimation of physical fatigue, cognitive load, and ergonomic risk in real manufacturing environments.
The kinematic-cognitive proxy hypothesis, that subtle degradation patterns detectable by CV serve as cognitive load indicators without brain sensors, requires validation in industrial settings.
Few-shot demographic adaptation, enabling calibration of a new worker's model within minutes of their shift, is a key methodological contribution.

\textbf{Direction~2 - Actionable XAI for Operational Explainability.}
Standard XAI methods produce technically valid but operationally inert outputs.
Role-specific, operator-facing explainability mechanisms are needed that translate model outputs into immediately actionable instructions:
not \textit{``variable X1 drives fatigue''} but \textit{``lower the bench 85\,cm to reduce overhead reaching.''} 
This requires formalizing the distinction between technical and operational explainability and validating role-tailored explanation formats through user studies with operators, safety managers, and occupational health practitioners.

\textbf{Direction~3 - Dynamic Bio-Scheduling in OM Optimizers.}
Existing human-aware OM models use analytical fatigue functions or static ergonomic constraints.
The key next step is the integration of real-time sensor-derived worker state estimates as dynamic objectives within scheduling and task allocation optimizers, replacing assumed functions with live physiological ground truth.
This requires formal modeling of the sensor-OM interface and
algorithmic development to handle the probabilistic, person-specific nature of physiological inputs within traditionally deterministic OM frameworks.

\textbf{Direction~4 - HITL Validation in Industrial Settings.}
The complete framework must be validated in a representative manufacturing scenario with real workers. Beyond tracking operational metrics, validation protocols must treat regional regulatory constraints as binding architectural benchmarks rather than compliance afterthoughts. Because physiological signals feeding Layer~1 constitute sensitive biometric data subject to strict privacy protections, and automated worker-monitoring tools carry high-risk regulatory classifications, field evaluations must formally test whether the Layer~2 pre-execution override and Layer~4's pre-deployment consultation steps genuinely safeguard worker autonomy.
Validation against expert-annotated ground truth (REBA/RULA, NASA-TLX, Borg RPE) in a controlled but ecologically valid industrial testbed is both necessary, to comprehensively measure ergonomic safety, operational efficiency, and trust calibration.

\textbf{Direction~5 - Humanoid Interaction as a Physical HITL Layer.}
The four-layer framework proposed in this survey treats the interaction layer (Layer~4) as primarily informational: workers receive explanations and provide feedback through digital interfaces. A natural extension of this architecture is to instantiate Layer~4 through a physically co-present humanoid agent. Rather than receiving an on-screen instruction to \textit{``lower the bench 85\,cm to reduce overhead reaching.''}, a worker could receive that guidance through a humanoid collaborator that demonstrates the adjustment, monitors compliance in real time, and adapts its communication strategy based on worker response. 

This configuration raises new research questions across all four layers: what physiological signals does a humanoid presence itself induce in the worker (Layer~1); how should OM decisions account for the cognitive and social load of humanoid co-presence (Layer~2); what explanation modalities are appropriate for an embodied agent (Layer~3); and how trust calibration dynamics shift when the adaptive agent possesses physical embodiment and anthropomorphic morphology (Layer~4). These questions define a critical research frontier that the present survey identifies for future exploration.
\section{Conclusion}

This paper has surveyed the state-of-the-art across three dimensions
critical to realizing the human-centric vision of I5.0: worker
state monitoring, human factors in Operations Management, and
Human-in-the-Loop architectures.
The survey reveals that while each dimension has produced significant
capabilities, their integration into a unified, sensor-driven,
worker-participatory system remains unachieved.

The central finding is that the technology to monitor worker
physiological and cognitive state in industrial settings is mature
enough to support operational integration, but the connection between
sensing and OM decision-making has not been established.
OM models incorporating human factors continue to rely on analytical
fatigue assumptions rather than real-time sensor data; HITL
architectures allowing workers to feedback on adaptive OM decisions remain largely absent from the manufacturing research reviewed here; and no system addresses
demographic diversity in the sensing layer.

\section*{Acknowledgment}
This research is sponsored by national funds through Fundação para a Ciência e a Tecnologia (FCT), under projects UID/00285/2025 (DOI: 10.54499/UID/00285/2025) and LA/P/0112/2020 (DOI: 10.54499/LA/P/0112/2020).

\bibliographystyle{IEEEtran}
\bibliography{ref.bib}
\vfill

\end{document}